\documentclass[twoside,slac_one]{revtex4}

\usepackage{xspace}
\usepackage{subfigure}
\usepackage{graphicx} 
\usepackage{url}
\usepackage{multirow}
\usepackage{fancyhdr}
\usepackage{amsmath}
\usepackage{bm}
\usepackage{amsxtra}
\usepackage{amssymb}
\usepackage{amsthm}
\usepackage{latexsym}
\usepackage{lscape}
\usepackage{hyperref}
\hypersetup{
  colorlinks=true,
  linkcolor=red,
  citecolor=red,
  urlcolor=blue
}

\input{latex-commands.sty}
\input{atlasphysics.sty}

\begin{document}

\title{Jet substructure in ATLAS}
\author{David W. Miller, on behalf of the ATLAS Collaboration}
\affiliation{SLAC National Accelerator Laboratory, Menlo Park, CA USA}
\altaffiliation{Now at The University Of Chicago, Enrico Fermi Institute.}

\begin{abstract}
Measurements are presented of the jet invariant mass and substructure in proton-proton collisions at $\sqrt{s} = 7$~TeV with the ATLAS detector using an integrated luminosity of 37\invpb. These results exercise the tools for distinguishing the signatures of new boosted massive particles in the hadronic final state. Two ``fat'' jet algorithms are used, along with the filtering jet grooming technique that was pioneered in ATLAS. New jet substructure observables are compared for the first time to data at the LHC. Finally, a sample of candidate boosted top quark events collected in the 2010 data is analyzed in detail for the jet substructure properties of hadronic ``top-jets'' in the final state. These measurements demonstrate not only our excellent understanding of QCD in a new energy regime but open the path to using complex jet substructure observables in the search for new physics.
\end{abstract}

\maketitle
\thispagestyle{fancy}


\graphicspath{{./}}

\section{Introduction}
\label{sec:intro}
The hadronic final state may be explored in terms of the structure and shape of the hadronic energy flow via the internal hard and soft structure of jets -- jet ``substructure.'' Such an approach seeks to measure important aspects of QCD and to identify patterns and topological features that are potentially indicative of the presence of new physics.  

Hadronic decays of heavy particles such as $W$ bosons~\cite{Seymour:1993mx, Butterworth:2002tt, cscnote}, top quarks~\cite{ATL-PHYS-PUB-2010-008, brooijmans2,brooijmans,Chekanov:2010vc,Chekanov:2010gv}, the Higgs boson~\cite{Butterworth:2008iy,ATLASHV}, and potential new particles~\cite{Butterworth:2009qa}, may be collimated into a single heavy jet characterized by distinct substructure and large mass. In this analysis, two ``fat'' jet algorithms, along with the filtering jet grooming technique developed in the search for a boosted Higgs boson, are used to measure the jet mass and substructure. Results are presented at the particle level and compared to three Monte Carlo simulation programs. A first measurement of the jet mass scale uncertainty in ATLAS is made, and consequences for searches based on jet mass are discussed. An additional jet substructure observable, the first \kt splitting scale or $\sqrt{d_{12}}$, is measured for the first time at the LHC. Finally, a sample of candidate boosted top quark events collected in the 2010 data is analyzed in detail for the jet substructure properties of hadronic ``top-jets'' in the final state.

\section{Data selection and Monte Carlo samples}
\label{sec:data}
The ATLAS detector~\cite{detPaper} provides nearly full solid angle coverage of the collision point with tracking detectors, calorimeters and muon chambers. Of these multiple subsystems the most relevant to the analyses presented here are the inner detector~\cite{InDetPerfPaper}, barrel and endcap calorimeters~\cite{Aad2010e, TileReadiness} and trigger~\footnote{ATLAS uses a right-handed coordinate system with its origin at the nominal interaction point in the center of the detector and the $z$-axis along the beam pipe. Cylindrical coordinates $(r,\phi)$ are used in the transverse plane, $\phi$ being the azimuthal angle around the beam pipe. Pseudorapidity and rapidity are defined as $\eta = 1/2 \times \mathrm{ln}((|p| + p_z)/(|p| - p_z))$ and $y = 1/2 \times \mathrm{ln}((E + p_z)/(E - p_z))$ respectively.}.

The entire 2010 dataset is used for the analysis of jet substructure, corresponding to an integrated luminosity of approximately $37\pm1.3$\invpb. The uncertainty of 3.4\% on the luminosity is obtained from the latest luminosity calibration using the full set of van der Meer scans performed in 2010~\cite{LumiCONF2011}.

A sample of high-\pT\ jets is selected via a single jet trigger with a nominal transverse energy threshold of 95~GeV at the electromagnetic scale. The offline selection requires that the highest \pT\ jet be within $|\eta| < 2.0$ and to have $\pt > 300$~GeV. Events are required to contain exactly one primary vertex in order to reduce the impact of mulitple simultanous proton-proton interactions, although a dedicated study demonstrating the impact of such pile-up is also presented. The \AKT jet algorithm~\cite{Cacciari:2008gp, Cacciari200657} used for most of the results presented yields regularly shaped jets that are well-suited for calibration. The Cambridge-Aachen (\CamKt) jet algorithm~\cite{Dokshitzer:1997in,Wobisch:1998wt} is also very appropriate for measurements of jet substructure in two-body decays such as that expected from a boosted Higgs decaying to two $b$ quarks. Jet radius parameters of $R=1.0$ and $R=1.2$ are used for the \AKT and \CamKt algorithms, respectively. Jet quality criteria such as the timing of calorimeters cells, the electromagnetic energy fraction, and pulse shape information are used to select only those events with well measured jets. These criteria are designed to remove events that are likely to have contamination due to beam-related backgrounds or detector defects. Since the primary vertex multiplicity is different between data and MC due to event pile-up, a re-weighting factor is applied to the MC prior to event selection. 

The data are compared to events simulated using two different approaches to MC event generation. The first uses direct perturbative calculation of the cross-section matrix elements in powers of the strong coupling constant, \alphas. For QCD jet production, this approach is used by \Alpgen~\cite{alpgen} and performed at leading-order (LO) in \alphas\ for each relevant partonic subprocess. The second approach implements a sampling of the phase-space available for gluon emission with some suitable approximations. The simulation programs \Pythia~\cite{Sjostrand2001} and \herwigpp~\cite{Bahr2008} both implement this approach for QCD jet production and use LO perturbative calculations of matrix elements for \TwoToTwo processes and rely on the parton shower to produce the equivalent of multi-parton final states. In the analysis presented here, we utilize \Alpgen~2.13 with exact matrix element calculations up to $n\leq6$ partons and interfaced to \Herwig~6.510~\cite{Corcella2001} to provide the parton shower and hadronization model and with \Jimmy~4.31~\cite{jimmy} for the underlying event model. Comparisons are also made to \Pythia~6.423 and \Herwigpp~2.4 which provide shower models that are $p^{2}_T$ ordered and angular ordered, respectively. Leading-order parton density functions are taken from the MRST2007 LO*~\cite{Martin:2009iq,Sherstnev:2007nd} set, unless stated otherwise.

The MC generated samples are passed through a full simulation of the ATLAS detector and trigger~\cite{simulation} based on GEANT4~\cite{geant4}. The Quark Gluon String (QGSP) model~\cite{QGS} is used for the fragmentation of the nucleus, and the Bertini cascade (BERT) model~\cite{Bertini} for the description of the interactions of the hadrons in the medium of the nucleus. Alternative GEANT4 physics lists, using a combination of the FRITIOF~\cite{Fritiof} and Bertini models and QGSP without Bertini are used as part of the studies to understand the uncertainties on the
jet energy and mass scale.

\section{Jet substructure}
\label{sec:results}
The jet mass and internal splitting scales are measured for a sample of inclusive jets with $\ptjet>300$~GeV~\cite{ATLAS-CONF-2011-073}. The two jet algorithms, \AKT and \CamKt, utilized for these measurements use wide characteristic radii ($R=1.0$ and $R=1.2$, respectively) as this improves efficiency in the case of searches for new hadronically decaying heavy boosted objects. Furthermore, the jet filtering procedure introduced in Refs.~\cite{Butterworth:2008iy,ATLASHV} is applied to the \CamKt algorithm and the resulting ``groomed'' jet mass is measured. 

Whereas the jet mass is the four-momentum sum of the clustered jet constituents, the \KT\ splitting scale is defined by first reclustering the constituents of the jet with the \KT\
algorithm~\cite{kt, kt2}. The final recombination step defines the splitting scale variable as

\begin{eqnarray}
  \sqrt{d_{12}}  &=& \mathrm{min}(p_{T,i},p_{T,j}) \times \delta R_{i,j}; \\
  \delta R_{i,j} &=& \sqrt{d\phi_{i,j}^2 + dy_{i,j}^2},
\end{eqnarray}

\noindent where $i, j$ represent the two proto-jets combined at the final step of the \kt\ algorithm. Due to the definition of the distance metric for \kt, the last recombination will often represent the two most widely separated and highest \pT jet constituents. Consequently, for a two-body heavy particle decay the final clustering step will usually be to combine those two decay products. The parameter $\sqrt{d_{12}}$ can therefore be used to distinguish heavy particle decays, which tend to be reasonably symmetric, from largely asymmetric QCD splittings. The expected value for a heavy particle decay is approximately $m/2$ whereas plain QCD jets will tend to have values $\lesssim 20 \GeV$ with a long continuous tail to high $\sqrt{d_{12}}$.

For \CamKt jets, the filtering jet grooming procedure aims to identify splittings in a jet which each contribute significantly to the jet invariant mass. The procedure is described in detail in Ref.~\cite{ATLAS-CONF-2011-073}. At each step during the \CamKt clustering the masses of the proto-jets to be combined are recorded. This clustering history is then searched in reverse for a step at which the jet mass falls by approximately one-third while the \pT\ of the proto-jets is roughly symmetric. The jet is then reclustered with a smaller jet radius (typically $R=0.3$), and if more than two subjets are found the three highest \pT\ subjets are retained and used to calculate the final jet kinematics.

The jet mass and splitting scale are corrected for detector detector effects via a bin-by-bin unfolding procedure. The final bin sizes are chosen such that bin migrations are small and the per-bin purity~\footnote{The purity here refers to the fraction of reconstructed jets in a bin which were also in the same bin in the truth distribution.} is greater than 50\%.

\begin{figure}
  \centering
  \subfigure[]{
    \includegraphics[width=0.47\columnwidth]{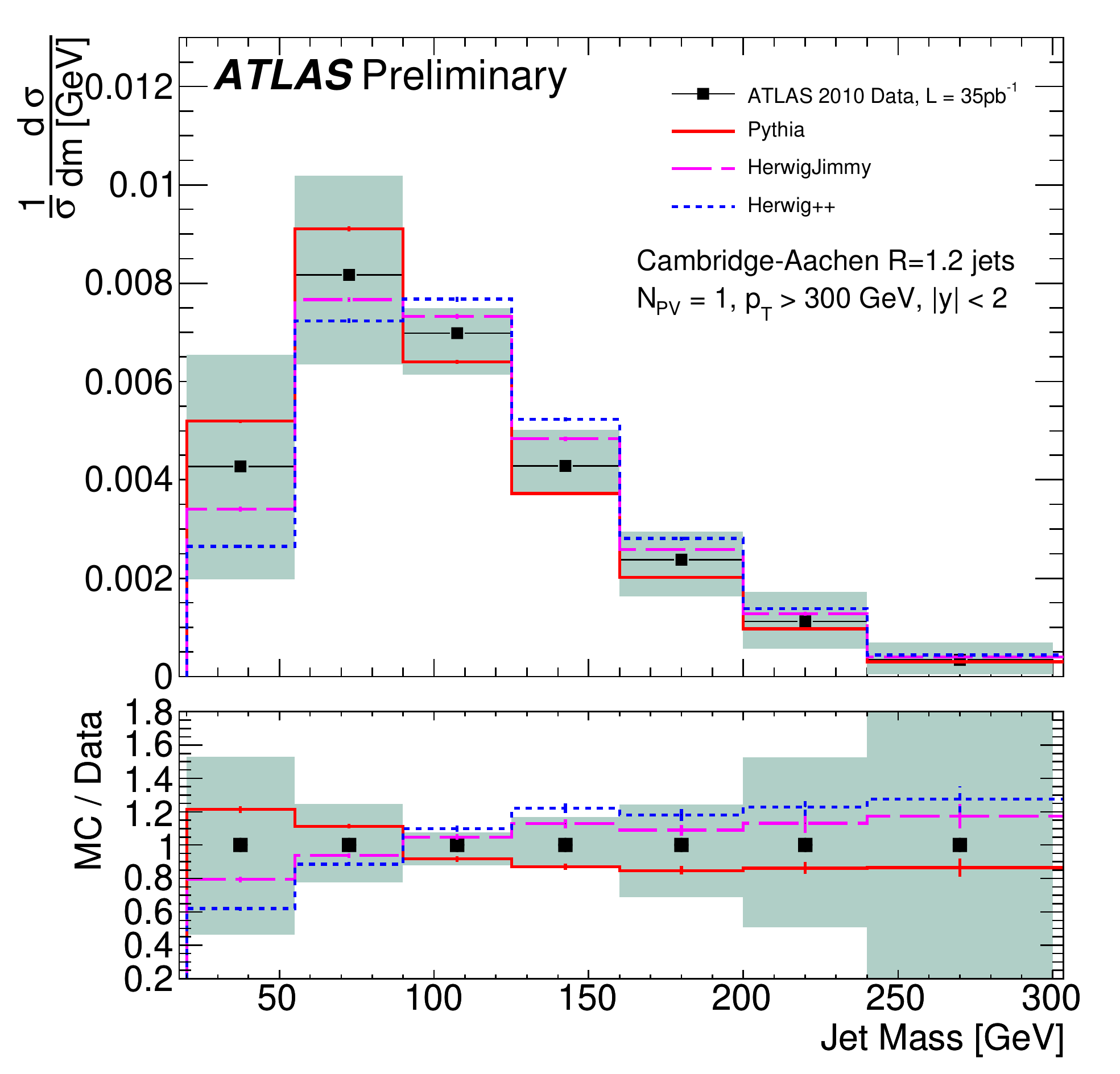}
    \label{fig:results:camass:unfiltered}}
  \subfigure[]{
    \includegraphics[width=0.47\columnwidth]{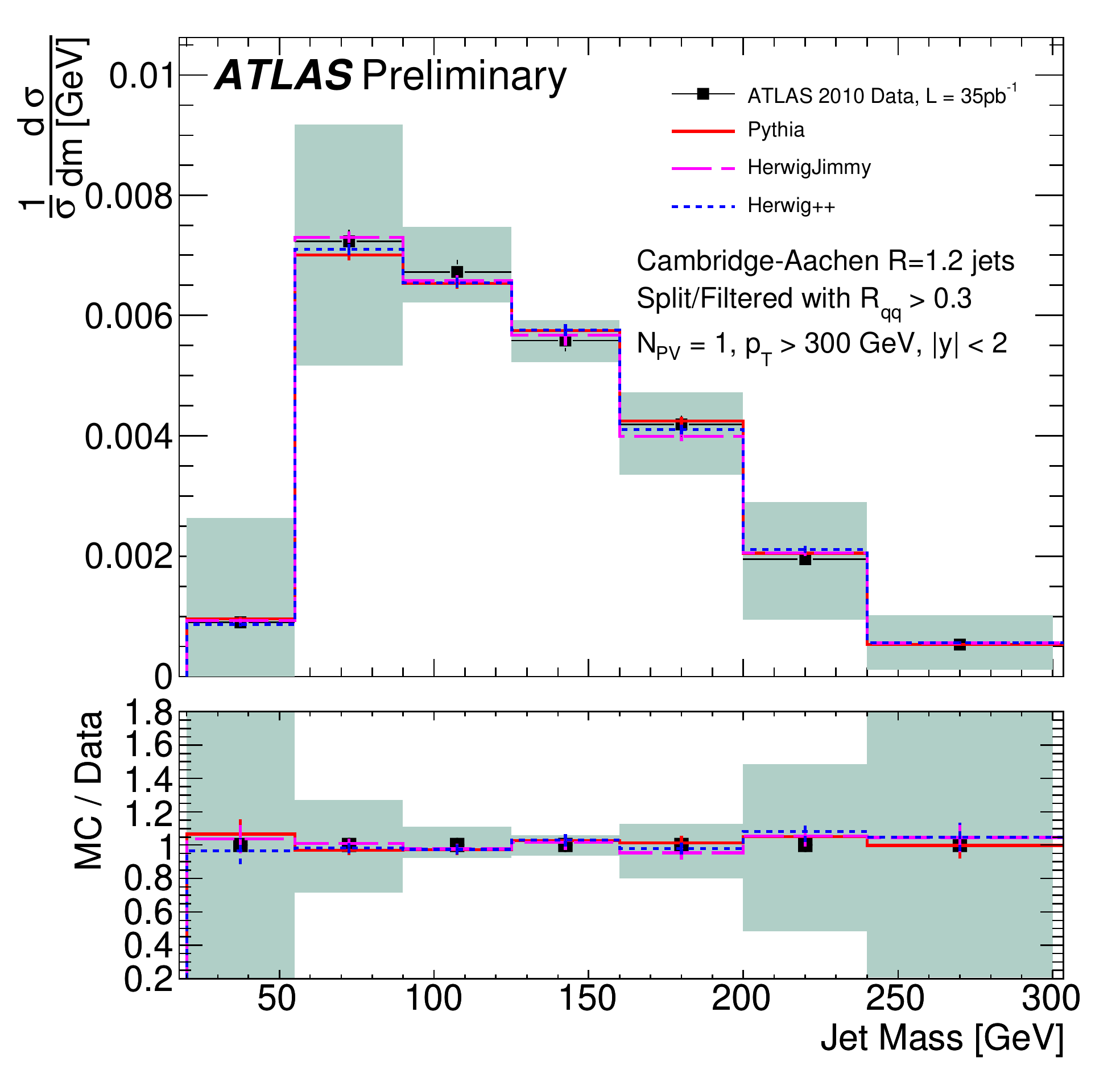}
    \label{fig:results:camass:filtered}}
  \caption{
    Invariant mass spectrum of Cambridge-Aachen jets with $p_{T} >
    300$~GeV and $|y| < 2$ (a) before 
    and (b) after the splitting and filtering
    procedure has been applied. Both distributions are fully corrected
    for detector effects, systematic uncertainties are depicted by the
    shaded band.
  }
  \label{fig:results:camass}
\end{figure}

\begin{figure}
  \centering
  \subfigure[]{
    \includegraphics[width=0.47\columnwidth]{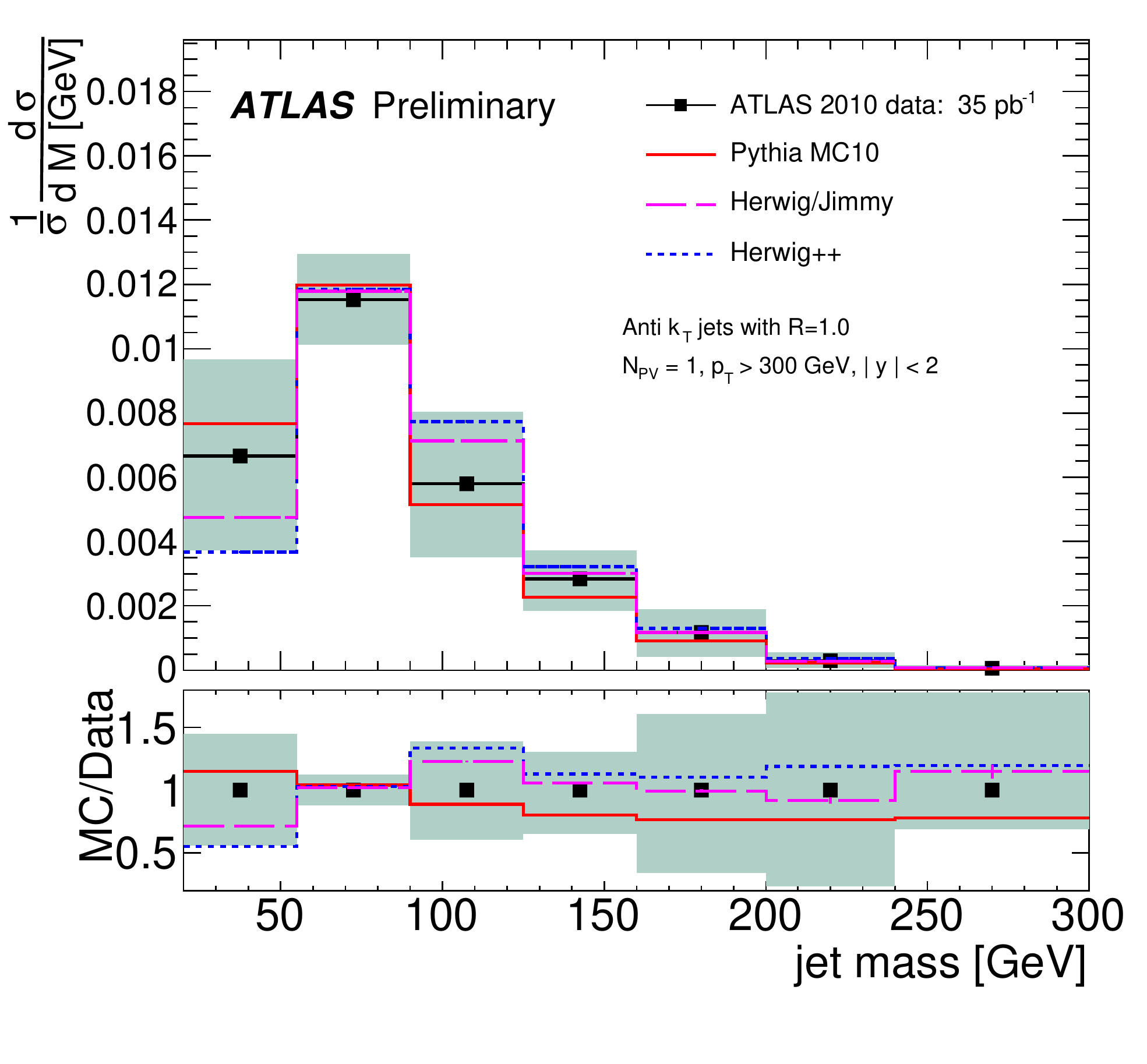}
    \label{fig:results:aktmass:mass}}
  \subfigure[]{
    \includegraphics[width=0.47\columnwidth]{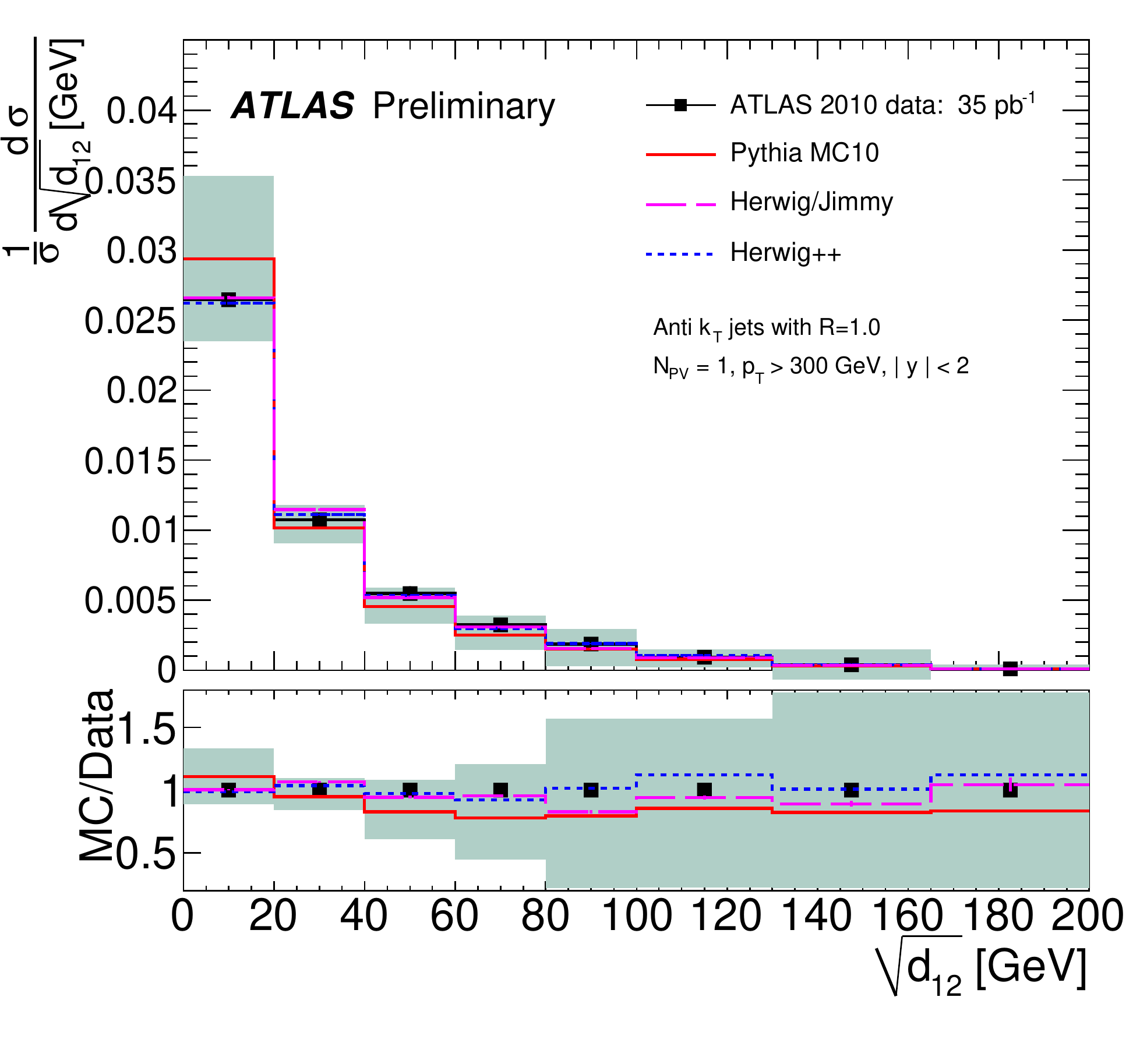}
    \label{fig:results:aktmass:d12}}
  \caption{
    (a) Invariant mass spectrum of \antikt jets with 
    $p_T > 300$~GeV and $|y| < 2$ and (b)
    $\sqrt{d_{12}}$ distribution for the same jets. Both distributions are fully 
    corrected for detector effects, systematic uncertainties are depicted by 
    the shaded band.
  }
  \label{fig:results:aktmass}
\end{figure}

Fully corrected hadron-level distributions of the jet mass and splitting scale are shown in Figures~\ref{fig:results:camass}-\ref{fig:results:aktmass}. Both the unfiltered \CamKt and filtered \CamKt jet mass distributions are shown in Figure~\ref{fig:results:camass}. \Herwigpp predicts slightly more massive jets than supported by the data whereas \Pythia and \HJ yield measurements which bracket the data and agree to within systematic uncertainties in all bins. Very notably, the differences between the mass distributions predicted by the various MC programs are greatly reduced after the filtering procedure as shown in Figure~\ref{fig:results:camass:unfiltered}. This observations suggests that the jet mass after the filtering procedure accurately represents the true hard components within the jets which are modeled well by the simulation.

Figure~\ref{fig:results:aktmass} depicts both the fully corrected mass distribution and the \kt\ splitting scale distribution for \AKTFat jets. A similar conclusion may be drawn in this case as in the \CamKt jets, wherein \Herwigpp tends to predict slightly more massive jets on average than observed in the data. However, in this case, the high mass tail of the distribution is actually better reflected in the \HJ and \Herwigpp MC simulations than in \Pythia. 

\begin{figure}
  \centering
  \subfigure[]{
    \includegraphics[width=0.47\columnwidth]{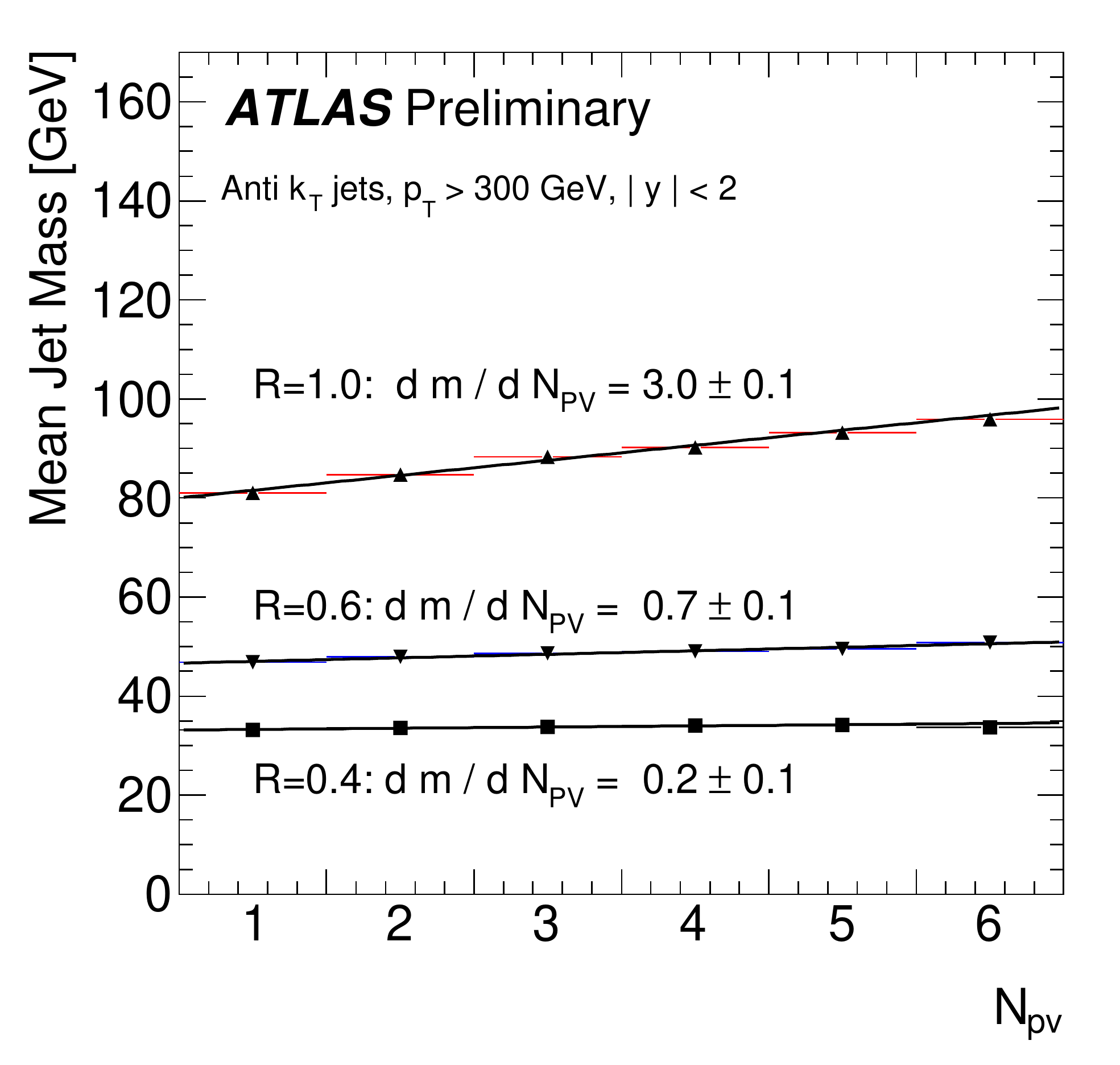}
    \label{fig:results:pileup:radius}}
  \subfigure[]{
    \includegraphics[width=0.47\columnwidth]{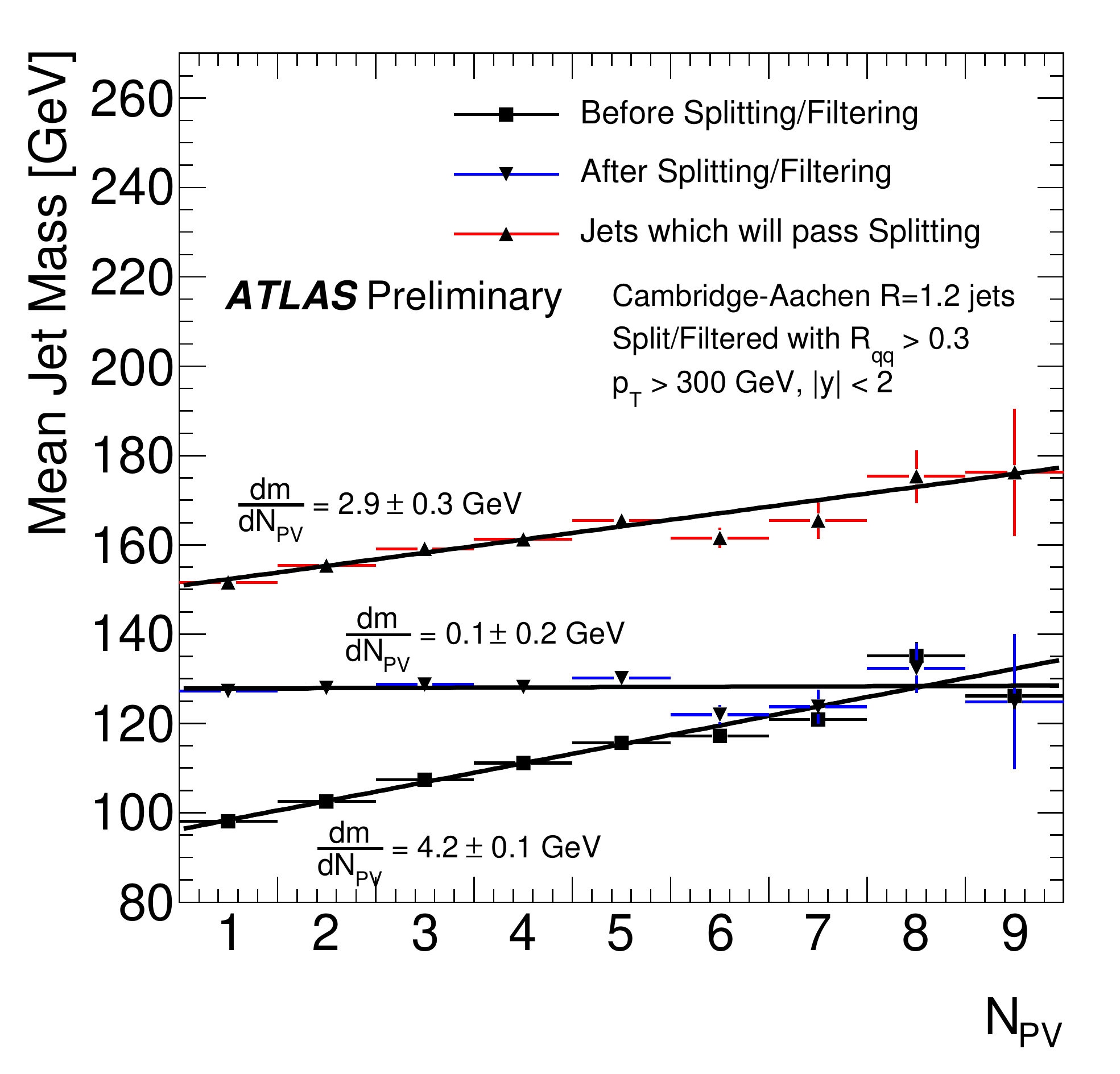}
    \label{fig:results:pileup:filtering}}
  \caption{
    The mean mass for jets with $p_{T} > 300$~GeV as a function of
    the number of primary vertices (\Npv) identified in the event.  Comparisons
    show the effect for \AKT jets with different resolution
    parameters (a) and \CAFat jets with and without splitting and
    filtering procedure (b). Each set of points is fitted with a
    straight line.
  }
  \label{fig:results:pileup}
\end{figure}

Figure~\ref{fig:results:pileup} exemplifies the potential for adverse pile-up effects and the power of grooming techniques such as jet filtering. The increase in the mean jet invariant mass of calorimeter jets is measured as a function of the number of additional interactions in the event, \Npv. Figure~\ref{fig:results:pileup:radius} highlights the dependence of this increase on the radius of the jet algorithm, growing linearly with $R$ in the case of zero pile-up, whereas the slope as a function of pile-up shows an $R^3$ dependence. This behavior can be qualitatively understood by the growth in jet area as $R^2$, and the contribution of additional particles to the jet mass scales with the distance between them $\simeq R/2$ yielding another power of $R$. Figure~\ref{fig:results:pileup:filtering} shows the rise in the mean jet mass as a function of \Npv before and after filtering. This procedure restores a flat dependence of the jet mass on \Npv to within uncertainties.

\section{Boosted heavy particles}
\label{sec:particles}
A handful of candidate boosted top quark events have been observed in the 2010 data and provide a unique first opportunity to test the techniques described above \textit{in-situ}.
This sample has been collected using the high purity lepton+jets channel, which has a signal-to-background (\StoB) ratio on the order 2 to 3~\cite{topxsec}. A subsample was then selected to have $m_{\ttbar}>700~\GeV$. These events offer a fertile testing ground for the substructure techniques discussed. 

\begin{figure}
  \centering
  \includegraphics[width=0.75\columnwidth]{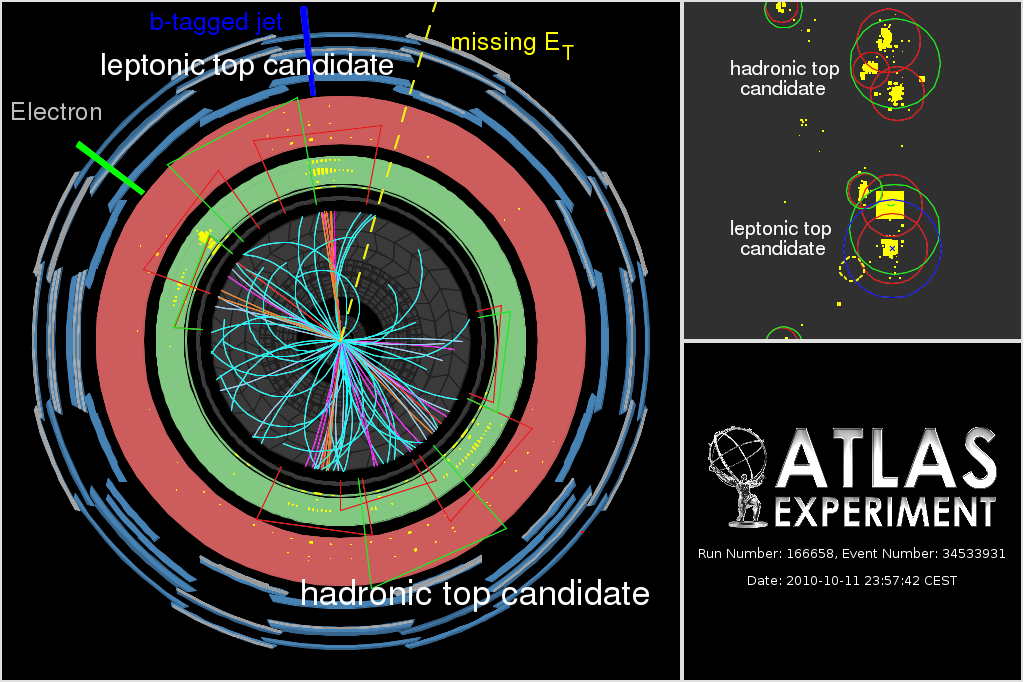}
  \caption{
     Leptonic top candidate formed by high $p_T$ electron (145~GeV),
     moderate $E_T^{miss}$, and a $b$-tagged jet. 
     When reclustered with $R=1.0$ the leptonic candidate acquires a large $p_T$, mass and 
     $1 \to 2 $ splitting scale as it absorbs the electron. Three jets 
     are identified with the hadronic top quark. When 
     reclustered with $R = 1.0$, these jets merge into a single jet with 
     $m_{\rm jet} = 197$~GeV, $\sqrt{d_{12}}=110$~GeV, and $\sqrt{d_{23}}=40$~GeV. 
     Jets indicated in red correspond to $R=0.4$, jets in green to $R=1.0$.
  }
  \label{fig:particles:top}
\end{figure}

Figure~\ref{fig:particles:top} shows a single ATLAS event display for one of the selected candidate boosted top quark events. In this display, \AKT jets with $R=0.4$ are indicated in red whereas large radius \AKT jets with $R=1.0$ are shown in green in the same figure. The leading $R=0.4$ jet in the event corresponds to the leptonic leg of the top event and has $\pT=199~\GeV$. It is also tagged as a $b$-jet with high probability via both a track impact parameter-based tagger and via a secondary-vertex + impact parameter tagger. The leading $R=0.4$ jet in the opposite hemisphere, corresponding to the hadronic leg of the event has $\pT=157~\GeV$ and no $b$-tag, whereas the second leading jet in the hemisphere of the hadronic candidate has a strong tag. When re-clustering the event with $R=1.0$, the jet corresponding to the hadronic leg has $\pT = 327~\GeV$ and $\mjet=206~\GeV$ as well as $\sqrt{d_{12}}=110$~GeV and $\sqrt{d_{23}}=40$~GeV. The value of $\sqrt{d_{23}}$ is suggestive of the presence of a boosted hadronically decaying $W$ boson reconstructed within this jet. These observations render this event to be a strong candidate for a boosted hadronically decaying top quark event.

\begin{figure}
  \centering
  \includegraphics[width=0.51\columnwidth]{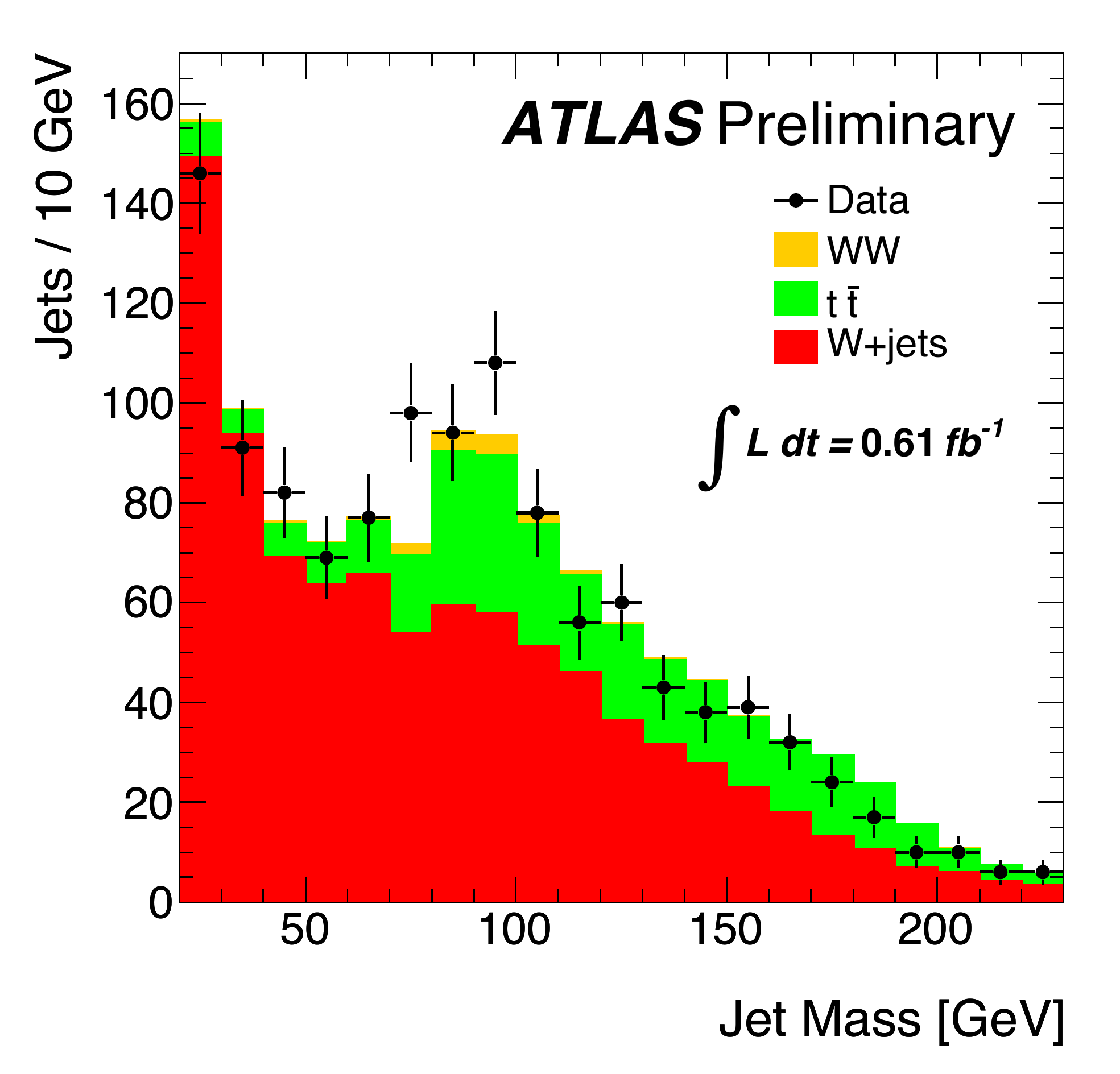}
  \caption{
    The jet mass distribution of subjets with $p_{T} > 180$~GeV in events 
    consistent with a $W \to l\nu$ boson decay with $p_{T} > 200$~GeV. 
    The distribution is compared to the uncorrected MC simulation prediction 
    for \ttbar, $W$+jets and $WW$ processes.
  }
  \label{fig:particles:Wmass}
\end{figure}

Lastly, a sample of events selected to contain a $W\to\ell\nu$ candidate consistent with having a $W$ boson $\pt^{W} > 200 \GeV$ is analyzed~\cite{ATLAS-CONF-2011-103}. The jet mass distribution of filtered \CAFat jets with $\pt > 180 \GeV$ and $\Delta \phi_{W,{\rm jet}} > 1.2$ in these events is shown in Figure~\ref{fig:particles:Wmass}. A clear mass peak near the $W$ mass indicates the presence of boosted hadronic $W$'s. The three main contributions to these events are \ttbar\ (generated with \MCATNLO\!+\HJ~\cite{mcatnlo,mcatnlo2}), $W$+jets (generated with \ApHJ),  and $WW$ (generated with \Herwig/\Jimmy), all normalized to the highest order cross-section available (see Ref.~\cite{ATLAS-CONF-2011-103} for more details). The good agreement between data and the various MC simulations suggests both that the tools described above are well described in a complex physics environment and that the systematics are generally well under control.

\section{Conclusions}
\label{sec:conclusions}
The study of the hadronic final state is explored in terms of the substructure of hadronic jets via measurements of the jet mass and \kt\ splitting scales for \AKT jets with $R=1.0$ and \CamKt jets with $R=1.2$. These measurements are corrected for detector effects and a full systematic uncertainty evaluation is performed. In addition, the jet mass is measured for \CamKt jets after a splitting and filtering procedure. In all observables the \Pythia and \Herwig samples are in agreement with data to within the systematic uncertainties. The \Herwigpp prediction appears to be slightly disfavoured in the unfiltered \CamKt mass spectra, producing jets with a higher mass than found in data.

The effect of pile-up on jet mass has also been studied. The expected scaling with number of vertices and jet radius is observed. It has also been shown that the filtering procedure applied to \CamKt jets reduces the impact of pile-up on jet mass to the extent that it is undetectable in the 2010 dataset. 

Finally, the applications of these techniques is demonstrated in the tagging of candidate boosted top quarks and the measurement of fully hadronic $W$ decays. 

Overall it is clear that ATLAS is capable of delivering measurements of the variables considered in this study and that these observables are well modeled by leading order Monte Carlo. Searches for boosted Higgs bosons, supersymmetric particles, and top-quark resonances will all benefit from utilizing these techniques.



\bigskip
\bibliography{substructure}

\begin{thebibliography}{39}
\expandafter\ifx\csname natexlab\endcsname\relax\def\natexlab#1{#1}\fi
\expandafter\ifx\csname bibnamefont\endcsname\relax
  \def\bibnamefont#1{#1}\fi
\expandafter\ifx\csname bibfnamefont\endcsname\relax
  \def\bibfnamefont#1{#1}\fi
\expandafter\ifx\csname citenamefont\endcsname\relax
  \def\citenamefont#1{#1}\fi
\expandafter\ifx\csname url\endcsname\relax
  \def\url#1{\texttt{#1}}\fi
\expandafter\ifx\csname urlprefix\endcsname\relax\def\urlprefix{URL }\fi
\providecommand{\bibinfo}[2]{#2}
\providecommand{\eprint}[2][]{\url{#2}}

\bibitem[{\citenamefont{Seymour}(1994)}]{Seymour:1993mx}
\bibinfo{author}{\bibfnamefont{M.~H.} \bibnamefont{Seymour}},
  \bibinfo{journal}{Z. Phys.} \textbf{\bibinfo{volume}{C62}},
  \bibinfo{pages}{127} (\bibinfo{year}{1994}).

\bibitem[{\citenamefont{Butterworth et~al.}(2002)\citenamefont{Butterworth,
  Cox, and Forshaw}}]{Butterworth:2002tt}
\bibinfo{author}{\bibfnamefont{J.~M.} \bibnamefont{Butterworth}},
  \bibinfo{author}{\bibfnamefont{B.~E.} \bibnamefont{Cox}}, \bibnamefont{and}
  \bibinfo{author}{\bibfnamefont{J.~R.} \bibnamefont{Forshaw}},
  \bibinfo{journal}{Phys. Rev.} \textbf{\bibinfo{volume}{D65}},
  \bibinfo{pages}{096014} (\bibinfo{year}{2002}), \eprint{hep-ph/0201098}.

\bibitem[{\citenamefont{{ATLAS Collaboration}}(2009{\natexlab{a}})}]{cscnote}
\bibinfo{author}{\bibnamefont{{ATLAS Collaboration}}}
  (\bibinfo{year}{2009}{\natexlab{a}}), \eprint{hep-ex/0901.0512}.

\bibitem[{\citenamefont{{ATLAS
  Collaboration}}(2010{\natexlab{a}})}]{ATL-PHYS-PUB-2010-008}
\bibinfo{author}{\bibnamefont{{ATLAS Collaboration}}},
  \textbf{\bibinfo{volume}{\href{http://cdsweb.cern.ch/record/1278454}{ATL-PHYS-PUB-2010-008}}}
  (\bibinfo{year}{2010}{\natexlab{a}}).

\bibitem[{\citenamefont{{ATLAS
  Collaboration}}(2009{\natexlab{b}})}]{brooijmans2}
\bibinfo{author}{\bibnamefont{{ATLAS Collaboration}}},
  \textbf{\bibinfo{volume}{\href{http://cdsweb.cern.ch/record/1177410}{ATL-PHYS-PUB-2009-081}}}
  (\bibinfo{year}{2009}{\natexlab{b}}).

\bibitem[{\citenamefont{{ATLAS
  Collaboration}}(2009{\natexlab{c}})}]{brooijmans}
\bibinfo{author}{\bibnamefont{{ATLAS Collaboration}}},
  \textbf{\bibinfo{volume}{\href{http://cdsweb.cern.ch/record/1077731}{ATL-PHYS-CONF-2008-008}}}
  (\bibinfo{year}{2009}{\natexlab{c}}).

\bibitem[{\citenamefont{Chekanov and Proudfoot}(2010)}]{Chekanov:2010vc}
\bibinfo{author}{\bibfnamefont{S.}~\bibnamefont{Chekanov}} \bibnamefont{and}
  \bibinfo{author}{\bibfnamefont{J.}~\bibnamefont{Proudfoot}},
  \bibinfo{journal}{Phys. Rev.} \textbf{\bibinfo{volume}{D81}},
  \bibinfo{pages}{114038} (\bibinfo{year}{2010}), \eprint{hep-ph/1002.3982}.

\bibitem[{\citenamefont{Chekanov et~al.}(2010)\citenamefont{Chekanov, Levy,
  Proudfoot, and Yoshida}}]{Chekanov:2010gv}
\bibinfo{author}{\bibfnamefont{S.}~\bibnamefont{Chekanov}},
  \bibinfo{author}{\bibfnamefont{C.}~\bibnamefont{Levy}},
  \bibinfo{author}{\bibfnamefont{J.}~\bibnamefont{Proudfoot}},
  \bibnamefont{and} \bibinfo{author}{\bibfnamefont{R.}~\bibnamefont{Yoshida}},
  \bibinfo{journal}{Phys. Rev.} \textbf{\bibinfo{volume}{D82}},
  \bibinfo{pages}{094029} (\bibinfo{year}{2010}), \eprint{hep-ph/1009.2749}.

\bibitem[{\citenamefont{Butterworth et~al.}(2008)}]{Butterworth:2008iy}
\bibinfo{author}{\bibfnamefont{J.~M.} \bibnamefont{Butterworth}}
  \bibnamefont{et~al.}, \bibinfo{journal}{Phys. Rev. Lett.}
  \textbf{\bibinfo{volume}{100}}, \bibinfo{pages}{242001}
  (\bibinfo{year}{2008}), \eprint{hep-ph/0802.2470}.

\bibitem[{\citenamefont{{ATLAS Collaboration}}(2009{\natexlab{d}})}]{ATLASHV}
\bibinfo{author}{\bibnamefont{{ATLAS Collaboration}}},
  \textbf{\bibinfo{volume}{\href{http://cdsweb.cern.ch/record/1201444}{ATL-PHYS-PUB-2009-088}}}
  (\bibinfo{year}{2009}{\natexlab{d}}).

\bibitem[{\citenamefont{Butterworth et~al.}(2009)}]{Butterworth:2009qa}
\bibinfo{author}{\bibfnamefont{J.~M.} \bibnamefont{Butterworth}}
  \bibnamefont{et~al.}, \bibinfo{journal}{Phys. Rev. Lett.}
  \textbf{\bibinfo{volume}{103}}, \bibinfo{pages}{241803}
  (\bibinfo{year}{2009}), \eprint{hep-ph/0906.0728}.

\bibitem[{\citenamefont{{ATLAS Collaboration}}(2008)}]{detPaper}
\bibinfo{author}{\bibnamefont{{ATLAS Collaboration}}}, \bibinfo{journal}{JINST}
  \textbf{\bibinfo{volume}{3}}, \bibinfo{pages}{S08003} (\bibinfo{year}{2008}).

\bibitem[{\citenamefont{{ATLAS
  Collaboration}}(2010{\natexlab{b}})}]{InDetPerfPaper}
\bibinfo{author}{\bibnamefont{{ATLAS Collaboration}}},
  \bibinfo{journal}{European Physical Journal C} \textbf{\bibinfo{volume}{70}},
  \bibinfo{pages}{787} (\bibinfo{year}{2010}{\natexlab{b}}),
  \eprint{physics.ins-det/1004.5293}.

\bibitem[{\citenamefont{{ATLAS Collaboration}}(2010{\natexlab{c}})}]{Aad2010e}
\bibinfo{author}{\bibnamefont{{ATLAS Collaboration}}},
  \bibinfo{journal}{European Physical Journal C} \textbf{\bibinfo{volume}{70}},
  \bibinfo{pages}{755} (\bibinfo{year}{2010}{\natexlab{c}}),
  \eprint{physics.ins-det/1002.4189}.

\bibitem[{\citenamefont{{ATLAS
  Collaboration}}(2010{\natexlab{d}})}]{TileReadiness}
\bibinfo{author}{\bibnamefont{{ATLAS Collaboration}}},
  \bibinfo{journal}{European Physical Journal C} \textbf{\bibinfo{volume}{70}},
  \bibinfo{pages}{1193} (\bibinfo{year}{2010}{\natexlab{d}}),
  \eprint{physics.ins-det/1007.5423}.

\bibitem[{\citenamefont{{ATLAS
  Collaboration}}(2011{\natexlab{a}})}]{LumiCONF2011}
\bibinfo{author}{\bibnamefont{{ATLAS Collaboration}}},
  \textbf{\bibinfo{volume}{\href{http://cdsweb.cern.ch/record/1334563}{ATLAS-CONF-2011-011}}}
  (\bibinfo{year}{2011}{\natexlab{a}}).

\bibitem[{\citenamefont{Cacciari et~al.}(2008)\citenamefont{Cacciari, Salam,
  and Soyez}}]{Cacciari:2008gp}
\bibinfo{author}{\bibfnamefont{M.}~\bibnamefont{Cacciari}},
  \bibinfo{author}{\bibfnamefont{G.~P.} \bibnamefont{Salam}}, \bibnamefont{and}
  \bibinfo{author}{\bibfnamefont{G.}~\bibnamefont{Soyez}},
  \bibinfo{journal}{JHEP} \textbf{\bibinfo{volume}{04}}, \bibinfo{pages}{063}
  (\bibinfo{year}{2008}), \eprint{0802.1189}.

\bibitem[{\citenamefont{Cacciari and Salam}(2006)}]{Cacciari200657}
\bibinfo{author}{\bibfnamefont{M.}~\bibnamefont{Cacciari}} \bibnamefont{and}
  \bibinfo{author}{\bibfnamefont{G.~P.} \bibnamefont{Salam}},
  \bibinfo{journal}{Phys. Lett. B} \textbf{\bibinfo{volume}{641}},
  \bibinfo{pages}{57 } (\bibinfo{year}{2006}), \eprint{hep-ph/0707.1378}.

\bibitem[{\citenamefont{Dokshitzer et~al.}(1997)}]{Dokshitzer:1997in}
\bibinfo{author}{\bibfnamefont{Y.~L.} \bibnamefont{Dokshitzer}}
  \bibnamefont{et~al.}, \bibinfo{journal}{JHEP} \textbf{\bibinfo{volume}{08}},
  \bibinfo{pages}{001} (\bibinfo{year}{1997}), \eprint{hep-ph/9707323}.

\bibitem[{\citenamefont{Wobisch and Wengler}(1998)}]{Wobisch:1998wt}
\bibinfo{author}{\bibfnamefont{M.}~\bibnamefont{Wobisch}} \bibnamefont{and}
  \bibinfo{author}{\bibfnamefont{T.}~\bibnamefont{Wengler}}
  (\bibinfo{year}{1998}), \eprint{hep-ph/9907280}.

\bibitem[{\citenamefont{Mangano et~al.}(2003)}]{alpgen}
\bibinfo{author}{\bibfnamefont{M.~L.} \bibnamefont{Mangano}}
  \bibnamefont{et~al.}, \bibinfo{journal}{JHEP} \textbf{\bibinfo{volume}{07}},
  \bibinfo{pages}{001} (\bibinfo{year}{2003}), \eprint{hep-ph/0206293}.

\bibitem[{\citenamefont{Sj\"{o}strand}(2001)}]{Sjostrand2001}
\bibinfo{author}{\bibfnamefont{T.}~\bibnamefont{Sj\"{o}strand}},
  \bibinfo{journal}{Computer Physics Communications}
  \textbf{\bibinfo{volume}{135}}, \bibinfo{pages}{238} (\bibinfo{year}{2001}),
  \eprint{hep-ph/0010017}.

\bibitem[{\citenamefont{B\"{a}hr et~al.}(2008)}]{Bahr2008}
\bibinfo{author}{\bibfnamefont{M.}~\bibnamefont{B\"{a}hr}}
  \bibnamefont{et~al.}, \bibinfo{journal}{European Physical Journal C}
  \textbf{\bibinfo{volume}{58}}, \bibinfo{pages}{639} (\bibinfo{year}{2008}),
  \eprint{{hep-ph/0803.0883}}.

\bibitem[{\citenamefont{Corcella et~al.}(2001)}]{Corcella2001}
\bibinfo{author}{\bibfnamefont{G.}~\bibnamefont{Corcella}}
  \bibnamefont{et~al.}, \bibinfo{journal}{Journal of High Energy Physics}
  \textbf{\bibinfo{volume}{2001}}, \bibinfo{pages}{010} (\bibinfo{year}{2001}),
  \eprint{hep-ph/0011363}.

\bibitem[{\citenamefont{Butterworth et~al.}(1996)\citenamefont{Butterworth,
  Forshaw, and Seymour}}]{jimmy}
\bibinfo{author}{\bibfnamefont{J.~M.} \bibnamefont{Butterworth}},
  \bibinfo{author}{\bibfnamefont{J.~R.} \bibnamefont{Forshaw}},
  \bibnamefont{and} \bibinfo{author}{\bibfnamefont{M.~H.}
  \bibnamefont{Seymour}}, \bibinfo{journal}{Z. Phys. C}
  \textbf{\bibinfo{volume}{72}}, \bibinfo{pages}{637} (\bibinfo{year}{1996}),
  \eprint{hep-ph/9601371}.

\bibitem[{\citenamefont{Martin et~al.}(2009)\citenamefont{Martin, Stirling,
  Thorne, and Watt}}]{Martin:2009iq}
\bibinfo{author}{\bibfnamefont{A.~D.} \bibnamefont{Martin}},
  \bibinfo{author}{\bibfnamefont{W.~J.} \bibnamefont{Stirling}},
  \bibinfo{author}{\bibfnamefont{R.~S.} \bibnamefont{Thorne}},
  \bibnamefont{and} \bibinfo{author}{\bibfnamefont{G.}~\bibnamefont{Watt}},
  \bibinfo{journal}{Eur. Phys. J.} \textbf{\bibinfo{volume}{C63}},
  \bibinfo{pages}{189} (\bibinfo{year}{2009}), \eprint{hep-ph/0901.0002}.

\bibitem[{\citenamefont{Sherstnev and Thorne}(2008)}]{Sherstnev:2007nd}
\bibinfo{author}{\bibfnamefont{A.}~\bibnamefont{Sherstnev}} \bibnamefont{and}
  \bibinfo{author}{\bibfnamefont{R.~S.} \bibnamefont{Thorne}},
  \bibinfo{journal}{Eur. Phys. J.} \textbf{\bibinfo{volume}{C55}},
  \bibinfo{pages}{553} (\bibinfo{year}{2008}), \eprint{hep-ph/0711.2473}.

\bibitem[{\citenamefont{{ATLAS
  Collaboration}}(2010{\natexlab{e}})}]{simulation}
\bibinfo{author}{\bibnamefont{{ATLAS Collaboration}}},
  \bibinfo{journal}{European Physical Journal C} \textbf{\bibinfo{volume}{70}},
  \bibinfo{pages}{823} (\bibinfo{year}{2010}{\natexlab{e}}).

\bibitem[{\citenamefont{Agostinelli et~al.}(2003)}]{geant4}
\bibinfo{author}{\bibfnamefont{S.}~\bibnamefont{Agostinelli}}
  \bibnamefont{et~al.} (\bibinfo{collaboration}{GEANT4}),
  \bibinfo{journal}{Nucl. Instrum. Meth. A} \textbf{\bibinfo{volume}{506}},
  \bibinfo{pages}{250} (\bibinfo{year}{2003}).

\bibitem[{\citenamefont{Folger and Wellisch}(2003)}]{QGS}
\bibinfo{author}{\bibfnamefont{G.}~\bibnamefont{Folger}} \bibnamefont{and}
  \bibinfo{author}{\bibfnamefont{J.~P.} \bibnamefont{Wellisch}}
  (\bibinfo{year}{2003}), \eprint{nucl-th/0306007}.

\bibitem[{\citenamefont{Bertini}(1969)}]{Bertini}
\bibinfo{author}{\bibfnamefont{H.~W.} \bibnamefont{Bertini}},
  \bibinfo{journal}{Phys. Rev. A} \textbf{\bibinfo{volume}{188}},
  \bibinfo{pages}{1711} (\bibinfo{year}{1969}).

\bibitem[{\citenamefont{Andersson et~al.}(1987)\citenamefont{Andersson,
  Gustafson, and Nilsson-Almqvist}}]{Fritiof}
\bibinfo{author}{\bibfnamefont{B.}~\bibnamefont{Andersson}},
  \bibinfo{author}{\bibfnamefont{G.}~\bibnamefont{Gustafson}},
  \bibnamefont{and}
  \bibinfo{author}{\bibfnamefont{B.}~\bibnamefont{Nilsson-Almqvist}},
  \bibinfo{journal}{Nucl. Phys. B} \textbf{\bibinfo{volume}{281}},
  \bibinfo{pages}{289 } (\bibinfo{year}{1987}).

\bibitem[{\citenamefont{{ATLAS
  Collaboration}}(2011{\natexlab{b}})}]{ATLAS-CONF-2011-073}
\bibinfo{author}{\bibnamefont{{ATLAS Collaboration}}},
  \textbf{\bibinfo{volume}{\href{http://cdsweb.cern.ch/record/1352454}{ATLAS-CONF-2011-073}}}
  (\bibinfo{year}{2011}{\natexlab{b}}).

\bibitem[{\citenamefont{Catani et~al.}(1993)\citenamefont{Catani, Dokshitzer,
  Seymour, and Webber}}]{kt}
\bibinfo{author}{\bibfnamefont{S.}~\bibnamefont{Catani}},
  \bibinfo{author}{\bibfnamefont{Y.~L.} \bibnamefont{Dokshitzer}},
  \bibinfo{author}{\bibfnamefont{M.~H.} \bibnamefont{Seymour}},
  \bibnamefont{and} \bibinfo{author}{\bibfnamefont{B.~R.}
  \bibnamefont{Webber}}, \bibinfo{journal}{Nucl. Phys.}
  \textbf{\bibinfo{volume}{B406}}, \bibinfo{pages}{187} (\bibinfo{year}{1993}).

\bibitem[{\citenamefont{Ellis and Soper}(1993)}]{kt2}
\bibinfo{author}{\bibfnamefont{S.~D.} \bibnamefont{Ellis}} \bibnamefont{and}
  \bibinfo{author}{\bibfnamefont{D.~E.} \bibnamefont{Soper}},
  \bibinfo{journal}{Phys. Rev.} \textbf{\bibinfo{volume}{D48}},
  \bibinfo{pages}{3160} (\bibinfo{year}{1993}), \eprint{hep-ph/9305266}.

\bibitem[{\citenamefont{{ATLAS Collaboration}}(2011{\natexlab{c}})}]{topxsec}
\bibinfo{author}{\bibnamefont{{ATLAS Collaboration}}},
  \bibinfo{journal}{European Physical Journal C} \textbf{\bibinfo{volume}{71}},
  \bibinfo{pages}{1577} (\bibinfo{year}{2011}{\natexlab{c}}),
  \eprint{hep-ex/1012.1792}.

\bibitem[{\citenamefont{{ATLAS
  Collaboration}}(2011{\natexlab{d}})}]{ATLAS-CONF-2011-103}
\bibinfo{author}{\bibnamefont{{ATLAS Collaboration}}},
  \textbf{\bibinfo{volume}{\href{http://cdsweb.cern.ch/record/1369826}{ATLAS-CONF-2011-103}}}
  (\bibinfo{year}{2011}{\natexlab{d}}).

\bibitem[{\citenamefont{Frixione and Webber}(2002)}]{mcatnlo}
\bibinfo{author}{\bibfnamefont{S.}~\bibnamefont{Frixione}} \bibnamefont{and}
  \bibinfo{author}{\bibfnamefont{B.~R.} \bibnamefont{Webber}},
  \bibinfo{journal}{JHEP} \textbf{\bibinfo{volume}{06}}, \bibinfo{pages}{029}
  (\bibinfo{year}{2002}), \eprint{hep-ph/0204244}.

\bibitem[{\citenamefont{Frixione et~al.}(2003)\citenamefont{Frixione, Nason,
  and Webber}}]{mcatnlo2}
\bibinfo{author}{\bibfnamefont{S.}~\bibnamefont{Frixione}},
  \bibinfo{author}{\bibfnamefont{P.}~\bibnamefont{Nason}}, \bibnamefont{and}
  \bibinfo{author}{\bibfnamefont{B.~R.} \bibnamefont{Webber}},
  \bibinfo{journal}{JHEP} \textbf{\bibinfo{volume}{08}}, \bibinfo{pages}{007}
  (\bibinfo{year}{2003}), \eprint{hep-ph/0305252}.

\end{thebibliography}

\end{document}